\documentclass[fleqn,12pt]{article}
\usepackage{epsfig}
\begin{document}
\textheight 22cm
\textwidth 15cm
\noindent
{\Large \bf Analytical theory of the probability distribution function of structure formation}
\newline
\newline
Johan Anderson\footnote{anderson.johan@gmail.com} and Eun-jin Kim
\newline
University of Sheffield
\newline
Department of Applied Mathematics
\newline
Hicks Building, Hounsfield Road 
\newline 
Sheffield
\newline
S3 7RH
\newline
UK
\newline
\newline
\begin{abstract}
\noindent
The probability distribution function (PDF) tails of the zonal flow structure formation and the PDF tails of momentum flux by incorporating effect of a shear flow in ion-temperature-gradient (ITG) turbulence are computed in the present paper. The bipolar vortex soliton (modon) is assumed to be the coherent structure responsible for bursty and intermittent events driving the PDF tails. It is found that stronger zonal flows are generated in ITG turbulence than Hasegawa-Mima (HM) turbulence as well as further from marginal stability. This suggests that although ITG turbulence has a higher level of heat flux, it also more likely generates stronger zonal flows, leading to a self-regulating system. It is also shown that shear flows can significantly reduce the PDF tails of Reynolds stress and structure formation. 
\end{abstract}
\newpage
\renewcommand{\thesection}{\Roman{section}}
\section{Introduction}
In order for magnetic confinement fusion energy to be a viable source of energy in our daily life, there are some current problems that have to be resolved. Among them is the control of the anomalous transport in Tokamaks, which is mostly attributed to a variety of micro-instabilities such as the ion-temperature-gradient (ITG) mode, trapped-electron (TE) mode and electron-temperature-gradient (ETG) mode~\cite{a10}. Micro-instabilities can however generate secondary modes e.g. the zonal flow that in turn may reduce the turbulent fluctuations in the plasma~\cite{a35}-~\cite{a36}. Note that zonal flows are poloidally and toroidally symmetric flow ($k_{\theta} =0$ and $k_{\parallel} = 0$) structures with a strong inhomogeneity in the radial direction ($k_r \neq 0$)~\cite{a51}-~\cite{a53}.

On the other hand coherent structures such as blobs and streamers can mediate significant heat transport through the formation of rare avalanche like events of large amplitude, as indicated by recent numerical studies~\cite{a14}-~\cite{a16}. Such events cause the deviation of the probability distribution functions (PDFs) from a Gaussian profile on which the traditional mean field theory (such as transport coefficients) is based. In particular, PDF tails due to rare events of large amplitude are often found to be substantially different from Gaussian although PDF centers tend to be Gaussian~\cite{a38}. These non-Gaussian PDF tails are manifestations of intermittency, caused by bursts and coherent structures. The characterization of these PDF tails thus requires a non-perturbative method. In previous works the PDF tails of momentum flux ($R$) and heat flux have been addressed~\cite{a19}-~\cite{a21}. It was shown that the PDF tail exhibits an ubiquitous exponential ($\simeq e^{-\xi R^{3/2}}$) and that the coefficient $\xi$ contains all the model dependent information, giving a difference between the ITG and the Hasegawa-Mima (HM) models~\cite{a81}; PDF tails in both cases significantly deviate from the Gaussian distributions for typical values of $\xi$. Moreover, it was found that reversed modon speed~\cite{a81} may have an important influence on the PDF tail~\cite{a42}.

In this paper we predict the PDF tails of the zonal flow structure formation and the PDF tails of momentum flux by incorporating the effect of a shear flow. The model used in the present study is an advanced fluid model for the ITG mode~\cite{a13} that permits the existence of a two dimensional bipolar vortex soliton (modon)~\cite{a81},~\cite{a25}-~\cite{a27}. Note that the ITG turbulence model has been successful in reproducing both experimental~\cite{a11} and non-linear gyro-kinetic results~\cite{a12}. The rare event of large amplitude represented by the modon can drive a zonal flow through the generalized Reynolds stress. The generation mechanism is studied in detail and the properties of the PDF tail of the zonal flow formation is studied. 

The theoretical technique used here is the so-called instanton method, a non-perturbative way of calculating the PDF tails. The PDF tail is first formally expressed in terms of a path integral by utilizing the Gaussian statistics of the forcing. An optimum path will then be associated with the creation of a modon (among all possible paths) and the action is evaluated using the saddle-point method on the effective action. The instanton is localized in time~\cite{a22}, existing during the formation of the modon. Thus, the bursty event can be associated with the creation of a modon.

It is found that the PDF tail of zonal flow ($\phi_{ZF}$) formation has a different exponential form ($\simeq e^{-\xi_2 \phi_{ZF}^{3}}$) compared with the momentum flux ($R$) and heat flux ($\simeq e^{-\xi_1 R^{3/2}}$) and that the constant $\xi_j$ ($j=1,2$) may also differ significantly. Furthermore the PDF tail of zonal flow formation in ITG turbulence is much stronger than that in HM turbulence. This suggests that stronger zonal flows are generated in ITG turbulence than in HM turbulence as well as further from marginal stability. Namely, while ITG turbulence is a state with high level of heat flux, it also generates strong zonal flows that inhibit transport. This suggests that turbulence and zonal flows constitute a self-regulating system whereby zonal flows generated by turbulence damp turbulence. It is shown that the shear flow can significantly reduce the PDF tails of Reynolds stress and the zonal flow formation.

The paper is organized as follows. In Sec. II the model of the ITG mode turbulence is presented together with preliminaries for the path-integral formulation for the PDF tails of structure formation. In Sec III the instanton solutions are calculated and the PDF tails of momentum flux and structure formation are estimated in Sec IV. We provide numerical results in Sec. V and a discussion of the results and conclusion in Sec. VI. 

\section{Non-perturbative calculation of structure formation PDF}
The ITG mode turbulence is modeled using the continuity and temperature equation for the ions and considering the electrons to be Boltzmann distributed~\cite{a13}. The effects of parallel ion motion, magnetic shear, trapped particles and finite beta on the ITG modes are neglected since in previous works the effect of parallel ion motion on the ITG mode was shown to be rather weak~\cite{a37}. The effect of a shear flow is incorporated in the time evolution equations for the background fluctuations in the form of sheared velocity $V_0$. The continuity and temperature equations then become,
\begin{eqnarray}
\frac{\partial n}{\partial t} - \left(\frac{\partial}{\partial t} - \alpha_i \frac{\partial}{\partial y}\right)\nabla^2_{\perp} \phi + \frac{\partial \phi}{\partial y} + V_0\frac{\partial}{\partial y} \left( 1 - \nabla_{\perp}^2 \right) \phi 
 - \nonumber \\  \epsilon_n  g_i \frac{\partial}{\partial y} \left(\phi + \tau \left(n + T_i \right) \right) + \nu \nabla^4 \phi =  - \left[\phi,n \right] + \left[\phi, \nabla^2_{\perp} \phi \right] + \tau \left[\phi, \nabla^2_{\perp} \left( n + T_i\right) \right]   + f\\
(\frac{\partial }{\partial t} + V_0 \frac{\partial}{\partial y})T_i - \frac{5}{3} \tau \epsilon_n g_i \frac{\partial T_i}{\partial y} + \left( \eta_i - \frac{2}{3}\right)\frac{\partial \phi}{\partial y} - \frac{2}{3} (\frac{\partial }{\partial t} + V_0 \frac{\partial}{\partial y}) n =  \nonumber \\
- \left[\phi,T_i \right] + \frac{2}{3} \left[\phi,n \right].
\end{eqnarray}
Eqs. (1)-(2) are closed by using the quasi-neutrality condition. Here $\left[ A ,B \right] = (\partial A/\partial x) (\partial B/\partial y) - (\partial A/\partial y) (\partial B/\partial x)$ is the Poisson bracket; $f$ is a forcing; $n = (L_n/\rho_s) \delta n / n_0$, $\phi = (L_n/\rho_s) e \delta \phi /T_e$, $T_i = (L_n/\rho_s) \delta T_i / T_{i0}$ are the normalized ion particle density, the electrostatic potential and the ion temperature, respectively. In equations (1) and (2), $\tau = T_i/T_e$, $\rho_s = c_s/\Omega_{ci}$ where $c_s=\sqrt{T_e/m_i}$, $\Omega_{ci} = eB/m_i c$ and $\nu$ is collisionality. We also define $L_f = - \left( d ln f / dr\right)^{-1}$ ($f = \{n, T_i \}$), $\eta_i = L_n / L_{T_i}$, $\epsilon_n = 2 L_n / \bar{R}$ where $\bar{R}$ is the major radius and $\alpha_i = \tau \left( 1 + \eta_i\right)$. The perpendicular length scale and time are normalized by $\rho_s$ and $L_n/c_s$, respectively. The geometrical quantities are calculated in the strong ballooning limit ($\theta = 0 $, $g_i \left(\theta = 0, \bar{\kappa} \right) = 1/\bar{\kappa}$  where $g_i \left( \theta \right)$ is defined by $\omega_D \left( \theta \right) = \omega_{\star} \epsilon_n g_i \left(\theta \right)$ and $\bar{\kappa}$ is the plasma cross-sectional elongation)~\cite{a28}-~\cite{a29}, with $\omega_{\star} = k_y v_{\star} = \rho_s c_s k_y/L_n $. It should be noted that the time evolution of the zonal flow potential ($\phi_{ZF}$) is governed by an averaged Eq. (1) [see last line in Eq. (6)]. This is equivalent to using the electron density response $n_e = \phi - \langle \phi \rangle$~\cite{a61}.

We calculate the PDF tails of momentum flux and zonal flow formation by using the instanton method. To this end, the PDF tail is expressed in terms of a path integral by utilizing the Gaussian statistics of the forcing $f$~\cite{a22}. The probability distribution function of Reynolds stress $Z_1 = R$ and zonal flow formation $Z_2 = \phi_{ZF}$ (using the notation $Z_j$ for j=\{1,2\}) can be defined as
\begin{eqnarray}
P(Z) & = &  \langle \delta(Z_j - Z) \rangle \nonumber \\
& = & \int d \lambda_j  \exp(i \lambda_j Z) \langle \exp(-i \lambda_j Z_j) \rangle \nonumber \\
& = & \int d\lambda_j \exp(i \lambda_j Z) I_{\lambda_j},
\end{eqnarray}
where 
\begin{eqnarray}
I_{\lambda_j} = \langle \exp(-i \lambda_j Z_j) \rangle.
\end{eqnarray}
The integrand can then be rewritten in the form of a path-integral as
\begin{eqnarray}
I_{\lambda_j} = \int \mathcal{D} \phi \mathcal{D} \bar{\phi} \mathcal{D} \phi_{ZF} \mathcal{D} \bar{\phi}_{ZF}  e^{-S_{\lambda_j}}.
\end{eqnarray}
Here, the parameter $j$ refers to the two specific cases included in the present study; namely $j=1$ gives the PDF tail of momentum flux while $j=2$ the PDF tail of the structure formation. The angular brackets denote the average over the statistics of the forcing $f$. By using the ansatz $T_i = \chi \phi$ [see Eq. (10)], the effective action $S_{\lambda_j}$ in Eq. (5) can be expressed as,
\begin{eqnarray}
S_{\lambda_j} & = & -i \int d^2x dt \bar{\phi} \left( \frac{\partial \phi}{\partial t} - (\frac{\partial }{\partial t} - \alpha_i \frac{\partial }{\partial y}) \nabla^2_{\perp} \phi + V_0(1-\nabla_{\perp}^2) \phi \right. \nonumber \\
& + & \left. (1-\epsilon_n g_i \beta)\frac{\partial \phi}{\partial y} - \beta [\phi, \nabla^2_{\perp} \phi ]\right) \nonumber \\
& + & \frac{1}{2} \int d^2x d^2 x^{\prime} \bar{\phi}(x) \kappa(x-x^{\prime}) \bar{\phi}(x^{\prime}) \nonumber \\
& + & i \lambda_1 \int d^2 x dt (-\frac{\partial \phi}{\partial x} \frac{\partial \phi}{\partial y}) \delta(t) \nonumber \\
& + & i \lambda_2 \int dt \phi_{ZF}(t) \delta(t) \nonumber \\
& - & i \int dt \bar{\phi}_{ZF}(t)(\frac{\partial \phi_{ZF}(t)}{\partial t} + \langle v_x v_y \rangle).
\end{eqnarray}
Note that the PDF tails of momentum flux and structure formation can be found by calculating the value of $S_{\lambda_j}$ at the saddle-point in the two cases $\lambda_1  \rightarrow \infty, \lambda_2 = 0$ or  $\lambda_1=0, \lambda_2  \rightarrow \infty$, respectively. Here the term $\langle v_x v_y \rangle$ is the Reynolds stress averaged over the forcing ($f$) and space. that generates the zonal flow. The first case $\lambda_1  \rightarrow \infty, \lambda_2 = 0$ gives the PDF tail of momentum flux in ITG turbulence including the interaction of a shear flow ($V_0$) while the second limit $\lambda_1=0, \lambda_2  \rightarrow \infty$ gives the PDF tail of zonal flow formation. In Eq. (6) we have used, 
\begin{eqnarray}
\beta & = & 1 + \tau + \tau \chi, \\
\chi & = & \frac{\eta_i - \frac{2}{3}(1-U + V_0)}{U - V_0 +\frac{5}{3}\tau \epsilon_n g_i}.
\end{eqnarray}

In Eq. (8), $U$ is the modon speed [see Eq. (10)]. To obtain Eq. (6) we have assumed the statistics of the forcing $f$ to be Gaussian with a short correlation time modeled by the delta function as
\begin{eqnarray}
\langle f(x, t) f(x^{\prime}, t^{\prime}) \rangle = \delta(t-t^{\prime})\kappa(x-x^{\prime}),
\end{eqnarray}
and $\langle f \rangle = 0$.
The delta correlation in time was chosen for the simplicity of the analysis. In the case of a finite correlation time the non-local integral equations in time are needed. We will also make use of the completeness of the Bessel function expansion and write $\kappa(x-x^{\prime}) = \kappa_0 (J_0(kx)J_0(kx^{\prime})+J_1(kx)J_1(kx^{\prime})(\cos \theta \cos \theta^{\prime} + \sin \theta \sin \theta^{\prime})+J_2(kx)J_2(kx^{\prime})(\cos 2 \theta \cos 2 \theta^{\prime} + \sin 2 \theta \sin 2 \theta^{\prime}) + ...)$.
\section{Instanton (saddle-point) solutions}
We have now reformulated the problem of calculating the PDF to a path-integral in Eq. (6). Although the path integral cannot in general be calculated exactly, an approximate value can be found in the limit $\lambda_j \rightarrow \infty$ by using a saddle point method. Since a direct application of the saddle-point equations results in very complicated partial differential equations for $\phi$ and $\bar{\phi}$, we assume that the instanton saddle-point solution is a temporally localized modon. Note that in Ref.~\cite{a81} the properties of the modons are explained in detail as well as some basic statistical models for an ideal gas of modons with the corresponding PDFs. That is, we assume that a short lived non-linear vortex soliton solution exists to the system of Eqs (1)-(2), by assuming that the electric potential $\phi$ and ion temperature $T_i$ can be written as
\begin{eqnarray}
\phi(x,y,t)& = &\psi(x,y-Ut)F(t), \mbox{and   } T_i = \chi \phi.
\end{eqnarray}
Here,
\begin{eqnarray}
\psi(x,y-Ut)& = & c_1 J_1(kr) (\cos \theta + \epsilon \sin \theta ) + \frac{\alpha}{k^2} r \cos \theta \mbox{   for   } r \leq a, \\
\psi(x,y-Ut)& = & c_2 K_1(pr) (\cos \theta + \bar{\epsilon}(r) \sin \theta )   \mbox{   for   } r \geq a. 
\end{eqnarray}
Here $J_1$ and $K_1$ are the first Bessel function and the second modified Bessel function, respectively; $r = \sqrt{x^2+y^2}$, $\tan \theta = y^{\prime}/x$, $y^{\prime} = y - Ut$, $\alpha = (A_1 - k^2 A_2)$, $A_1 = (1-\epsilon_n g_i - U + V_0)/\beta$, $A_2 = (U+\alpha_i - V_0)/\beta$. By matching the inner and outer solution at $r=a$ we find the conditions $c_1 = -\alpha a/J_1(ka)$, $c_2 = -Ua/K_1(pa)$, $J_1^{\prime}(ka)/J_1(ka) = (1+k^2/p^2)/ka - kK_1^{\prime}(pa)/pK_1(pa)$; $U$ is the velocity of the modon, and $a$ is the size of the core region. The function $\bar{\epsilon}(r)$ is chosen such that the matching conditions are similar to those in previous previous studies~\cite{a19}-~\cite{a20}. It is important to note that when $\epsilon = 0$ the Reynolds stress vanishes. The action $S_{\lambda_j}$ is then to be expressed only as an integral in time by using the conjugate variables,
\begin{eqnarray}
\bar{F}_0 & = & \int d^2 x \bar{\phi}(x,t) J_0(kr), \\
\bar{F}_{1s} & = & \int d^2 x \bar{\phi}(x,y,t) J_1(kr) \sin \theta, \\
\bar{F}_{1c} & = & \int d^2 x \bar{\phi}(x,y,t) J_1(kr) \cos \theta, \\
\bar{F}_{2s} & = & \int d^2 x \bar{\phi}(x,y,t) J_2(kr) \sin 2 \theta, \\
\bar{F}_{2c} & = & \int d^2 x \bar{\phi}(x,y,t) J_2(kr) \cos 2 \theta.
\end{eqnarray}
Note that the contribution from the outer solution ($r>a$) to $S_{\lambda_j}$ is neglected compared to that from the inner solution ($r<a$) for simplicity. This can be justified since the outer solution decays fast and inherently gives a minor contribution to the PDF tail. The action $S_{\lambda_j}$ consists of four different parts; the ITG model, the forcing, the Reynolds-stress parts or the structure formation and the time evolution of the zonal flow respectively. We note that the spatial structure of the zonal flow is not given which would only influence the constant $\phi_{ZF}(0)$ in Eq. (36), acting as a normalization in Eq. (64). The full action including the forcing and Reynolds stress or structure formation terms can then be expressed in terms of $F$, $\dot{F}$ and the conjugate variables $\bar{F}$,
\begin{eqnarray}
S_{\lambda_j} & = & -i \int dt [\gamma_1 \dot{F}(\bar{F}_{1c} + \epsilon \bar{F}_{1s})+  F(\gamma_2 \bar{F}_{2s} + \epsilon \gamma_3 \bar{F}_{0} + \epsilon \gamma_4 \bar{F}_{2c}) \nonumber \\
& + & F^2(\gamma_5 \bar{F}_{2s} + \epsilon \gamma_6 \bar{F}_{0} + \epsilon \gamma_7\bar{F}_{2c}) +  \gamma_8 (\bar{F}_{1c} + \epsilon \bar{F}_{1s}) )] \nonumber \\
& + & \frac{1}{2} \kappa_0 \int dt (\bar{F}_{0}^2 + 2(\bar{F}_{1c}^2 + \bar{F}_{1s}^2) + 2(\bar{F}_{2s}^2 + \bar{F}_{2c}^2)) \nonumber \\& + &  i \lambda_1 R_0 \int dt F^2(t) \delta(t) \nonumber \\
& + &  i \lambda_2 \int dt \phi_{ZF}  \delta(t) \nonumber \\
& - & i \int dt \bar{\phi}_{ZF}(\frac{\partial \phi_{ZF}}{\partial t} + \bar{R}_0 F^2)
\end{eqnarray}
Here the coefficients are,
\begin{eqnarray}
\gamma_1 & = & c_1(1 + k^2 + \frac{2 \alpha}{k^3}), \\
\gamma_2 & = & -\frac{k}{2} \alpha_1 = - \gamma_3 = - \frac{1}{2} \gamma_4, \\
\gamma_5 & = & -\frac{k}{2} \beta \alpha = - \gamma_6, \\
\gamma_7 & = & \beta k \alpha, \\
\gamma_8 & = & \nu k^4, \\
\alpha_1 & = & 1 - \epsilon_n g_i\beta - U - k^2(U+\alpha_i) + V_0(1+k^2) \\
R_0 & = & \int d^2 x (-\frac{\partial \psi}{\partial x} \frac{\partial \psi}{\partial y}) \\
\bar{R}_0 & = & \frac{R_0}{\pi a^2}.
\end{eqnarray}
Note that the effect of the shear flow ($V_0$) is incorporated in the parameter $\alpha$ and $\alpha_1$.

The equation of motion for the instanton is found by the variation of the action with respect to $F$, $\bar{F}_{0}$, $\bar{F}_{1c}$, $\bar{F}_{1s}$, $\bar{F}_{2c}$ and $\bar{F}_{2s}$,
\begin{eqnarray}
\frac{\delta S_{\lambda_1}}{\delta F} & = & -i[-\gamma_1 (\dot{\bar{F}}_{1c} + \epsilon \dot{\bar{F}}_{1s}) + (\gamma_2 \bar{F}_{2s} + \epsilon \gamma_3 \bar{F}_{0} + \epsilon \gamma_4 \bar{F}_{2c}) \nonumber \\
& + & 2F (\gamma_5 \bar{F}_{2s} + \epsilon \gamma_6 \bar{F}_{0} + \epsilon \gamma_7\bar{F}_{2c}) +  \gamma_8 (\bar{F}_{1c} + \epsilon \bar{F}_{1s}) ] \nonumber \\
& - &  2 i \bar{R}_0 \bar{\phi}_{ZF} F  - 2 i \lambda_1 R_0 F \delta(t) = 0, \\
\frac{\delta S_{\lambda_2}}{\delta F} & = & -i[-\gamma_1 (\dot{\bar{F}}_{1c} + \epsilon \dot{\bar{F}}_{1s}) + (\gamma_2 \bar{F}_{2s} + \epsilon \gamma_3 \bar{F}_{0} + \epsilon \gamma_4 \bar{F}_{2c}) \nonumber \\
& + & 2F (\gamma_5 \bar{F}_{2s} + \epsilon \gamma_6 \bar{F}_{0} + \epsilon \gamma_7\bar{F}_{2c}) +  \gamma_8 (\bar{F}_{1c} + \epsilon \bar{F}_{1s}) ] \nonumber \\
& - & 2 i \bar{R}_0 \bar{\phi}_{ZF} F = 0, \\
\frac{\delta S_{\lambda_j}}{\delta \bar{F}_0} & = & -i\epsilon ( \gamma_3 F + \gamma_6 F^2) + \kappa_0 \bar{F}_0 = 0, \\
\frac{\delta S_{\lambda_j}}{\delta \bar{F}_{1c}} & = & -i(\gamma_1 \dot{F} + \gamma_8 F) + 2 \kappa_0 \bar{F}_{1c} = 0, \\
\frac{\delta S_{\lambda_j}}{\delta \bar{F}_{1s}} & = & -i\epsilon (\dot{F} + \gamma_8 F) + 2 \kappa_0 \bar{F}_{1s} = 0, \\
\frac{\delta S_{\lambda_j}}{\delta \bar{F}_{2c}} & = & -i\epsilon (\gamma_4 F + \gamma_7 F^2) + 2 \kappa_0 \bar{F}_{2c} = 0, \\
\frac{\delta S_{\lambda_j}}{\delta \bar{F}_{2s}} & = & -i (\gamma_2 F + \gamma_5 F^2) + 2 \kappa_0 \bar{F}_{2s} = 0, \\
\frac{\delta S_{\lambda_j}}{\delta \bar{\phi}_{ZF}} & = & \dot{\phi}_{ZF} + \bar{R}_0 F^2 = 0, \\
\frac{\delta S_{\lambda_1}}{\delta \phi_{ZF}} & = & \dot{\bar{\phi}}_{ZF} = 0, \\
\frac{\delta S_{\lambda_2}}{\delta \phi_{ZF}} & = & \dot{\bar{\phi}}_{ZF} + \lambda_2 \phi_{ZF 0} \delta(t) = 0.
\end{eqnarray}
Here the initial condition for the zonal flow is used as $\phi_{ZF 0} = \phi_{ZF}(0)$. The equation of motion for $F$ is derived for $t < 0$ using Eqs. (27)-(36) as,
\begin{eqnarray}
\frac{1}{2} \gamma_1^2(1 + \epsilon^2 ) \frac{d \dot{F}^2}{dF}& = & \eta_1 F + 3 \eta_2 F^2 + 2 \eta_3 F^3 - 4 \kappa_0 \bar{\phi}_{ZF} F,
\end{eqnarray}
where
\begin{eqnarray}
\eta_1 & = & \gamma_2^2 + 2 \epsilon^2 \gamma_3^2 + \epsilon^2 \gamma_4^2 + \gamma_8^2 + \epsilon^2 \gamma_8^2, \\
\eta_2 & = & \gamma_2 \gamma_5 + 2 \epsilon^2 \gamma_3 \gamma_6 + \epsilon^2 \gamma_4 \gamma_7, \\
\eta_3 & = & \gamma_5^2 + 2 \epsilon^2 \gamma_6^2 + \epsilon^2 \gamma_7^2. 
\end{eqnarray}
Ordering the terms on the RHS of Eq. (37) in powers of $\lambda_2$ gives that the dominating terms are $2 \eta_3 F^3$ and $4 \kappa_0 \bar{\phi}_{ZF} F$ in the structure formation limit ($\lambda_2 \rightarrow \infty$ and $\lambda_1 = 0$). The contribution from the dissipation ($\nu$) to the term involving the time derivative of $F$ ($\dot{F}$) cancels out and the equation of motion is exactly solvable. In the limits of $\lambda_2 \rightarrow \infty$ and $\lambda_1 = 0$, Eqs (27)-(36) give us,
\begin{eqnarray}
\dot{F} & \simeq & F \sqrt{AF^2 + C}, \\
A & = & \frac{\eta_3}{(1+\epsilon^2) \gamma_1^2}, \\ 
C & = &  \frac{2 \kappa_0 \lambda_2}{(1+\epsilon^2) \gamma_1^2}.
\end{eqnarray}
Separation of variables then leads to,
\begin{eqnarray}
\sqrt{A}t = \frac{1}{2 \sqrt{C_0}} \ln \left(\frac{\sqrt{F^2+C_0} - \sqrt{C_0}}{\sqrt{F^2+C_0} + \sqrt{C_0}} \frac{\sqrt{F_0^2+C_0} - \sqrt{C_0}}{\sqrt{F_0^2+C_0} - \sqrt{C_0}} \right).
\end{eqnarray}
Here $C_0 = C/A$. The general solution to Eq. (44) can be written in terms of $H$ as
\begin{eqnarray}
H(t) & = & H_0 e^{\pm \sqrt{C} t},
\end{eqnarray}
where
\begin{eqnarray}
H(t) & = & \frac{\sqrt{F^2+C_0} - \sqrt{C_0}}{\sqrt{F^2+C_0} + \sqrt{C_0}}, \\
H_0 & = & \frac{\sqrt{F_0^2+C_0} - \sqrt{C_0}}{\sqrt{F_0^2+C_0} - \sqrt{C_0}}.
\end{eqnarray}
The time dependent function $F$ can now be determined by using Eq. (45) - (47),
\begin{eqnarray}
F(t) & = & \pm \frac{2 \sqrt{C_0 H_0}e^{\pm \sqrt{C}t}}{1-H_0 e^{2 \sqrt{C}t}}, \\
H_0 & = & 4 A - 2. 
\end{eqnarray}
Eq. (48) is derived from Eq. (44) by solving for F(t) and substituting H(t) using Eq. (45). The factor that determines the localization of $F(t)$ in time is given by the initial condition for the conjugate variable of the zonal flow. The initial condition is found by integrating Eq. (36) over the interval $[-\delta, 0]$ ($\delta \ll 1 $) and observing that the conjugate variables mediating between the forcing and $F$ vanish for $t \geq 0$. The initial condition for the zonal flow can be written,
\begin{eqnarray}
\bar{\phi}_{ZF}(-\delta) + \lambda_2 \phi_{ZF 0} = 0. 
\end{eqnarray}
For $t < 0$, we have $\dot{\bar{\phi}}_{ZF} = 0$.
We use this initial condition to compute the saddle-point action and then predict the scaling of $S_{\lambda_2}$ (as $\lambda_2 \rightarrow \infty$) to compute the PDF tail in the next section. The limit where $\lambda_1 \rightarrow \infty$ and $\lambda_1 = 0$ is treated in a similar way as in Ref.~\cite{a42} to compute the PDF tails of momentum flux. Following the steps in Eq. (37) - (50), but using a different $C$,

\begin{eqnarray}
C  =  \frac{2 \kappa_0 \lambda_1^2}{(1+\epsilon^2) \gamma_1^2},
\end{eqnarray}
we can determine the initial condition for the modon (F(0)) for the PDF tail of momentum flux. The initial condition for the zonal flow in this case is found from Eq. (38) or $\dot{\bar{\phi}}_{ZF} = 0$. Here the zonal flow and the conjugate zonal flow potential ($\phi_{ZF}$ and $\bar{\phi}_{ZF}$) is evolving passively following the drift wave and the conjugate drift wave potential ($\phi$ and $\bar{\phi}$).

\section{The PDF tails}
The PDF tails are found by calculating the value of $S_{\lambda_j}$ using Eq. (18) at the saddle-point in the two cases; the PDF tail of momentum flux by taking into account the effect of a shear flow ($\lambda_1 \rightarrow \infty, \lambda_2 = 0$) and the PDF tail of structure formation of zonal flow ($\lambda_1 = 0, \lambda_2 \rightarrow \infty$). The integral in Eq (18) is divided in four parts $C_1$ the ITG integral; $C_2$ the forcing integral; $C_3$ the momentum flux integral; $C_4$ the zonal flow structure integral and finally $C_5$ represents the zonal flow evolution integral. The resulting integrals in the first case are;
\begin{eqnarray}
S_{\lambda_1} & \simeq & - \frac{1}{3}h \lambda_1^3, \\
h & = & C_1 + C_2 + C_3 + C_4 + C_5, \\
C_1 & = & \frac{1}{2\kappa_0} \left( \gamma_1^2(1+\epsilon^2) [(\frac{4H_0}{H_0-1}-1)^{3/2}-1] \frac{C^{3/2}}{A} \right. \\
& +  &\left. 24 (\gamma_5^2 + 2 \epsilon^2 \gamma_6^2+\epsilon^2 \gamma_7^2) \frac{C^{3/2}}{A^{5/2}} (\frac{1}{3} \frac{H_0}{(H_0-1)^3} - \frac{1}{4} \frac{H_0}{(H_0-1)^2}) \right), \\
C_2 & = & \frac{1}{2\kappa_0} \left( \gamma_1^2(1+\epsilon^2) [(\frac{4H_0}{H_0-1}-1)^{3/2}-1] \frac{C^{3/2}}{2A} \right. \\
& +  &\left. 24 (\frac{1}{2}\gamma_5^2 + 2 \epsilon^2 \gamma_6^2) \frac{C^{3/2}}{A^{5/2}} (\frac{1}{3} \frac{H_0}{(H_0-1)^3} - \frac{1}{4} \frac{H_0}{(H_0-1)^2}) \right), \\
C_3 & = & R_0 F^2(0), \\ 
C_5 & = & \frac{2R_0 \sqrt{C}}{A} \frac{1}{H_0 - 1}, \\
H_0 & = & 4A-2, \\
A & = & \frac{\eta_3}{(1+\epsilon^2)\gamma_1^2}.
\end{eqnarray}
Note that here $C_4$ vanishes. In determining the integrals in Eq. (18) we have used the localized modon solution in Eq. (48). By using C defined in Eq. (51) and A in Eq. (61), we can find the coefficient $h$ in terms of the modon parameters using Eqs (19) - (25). The PDF tail of the Reynolds stress ($R$) can now be found by performing the integration over $\lambda_1$ in Eq. (6) using the saddle-point method in the same fashion as done in Ref.~\cite{a42} i.e. recall Eq. (6) gives $P(R) \sim \int d \lambda_1 \exp\{-\lambda_1 R - S_{\lambda_1}\} \sim \int d \lambda_1 \exp \{ - \lambda_1 R + h \lambda_1^3\}$. Now, the saddle point integral is evaluated at the maximum point $\lambda_{1 MAX} = \sqrt{R/(3 h)}$ with the result
\begin{eqnarray}
P(R) & \sim & \exp\{-\xi_1 (\frac{R}{R_0})^{3/2}\}, \\
\xi_1 & = &  \frac{2}{3}\frac{1}{\sqrt{3h}}.
\end{eqnarray}
This result is similar to the previous results where a similar exponential PDF was found. However, here the coefficient $\xi_1$ is modified by the presence of the shear flow ($V_0$).

In the second limit where $\lambda_1 = 0 \ \mbox{and} \ \lambda_2 \rightarrow \infty$ gives the PDF tail of the structure formation itself, the action integral yields
\begin{eqnarray}
S_{\lambda_2} & \simeq & - \frac{1}{3}h \lambda_2^{3/2}, \\
C_4 & = & 2 \phi_{ZF 0} \frac{\sqrt{C}}{A}.
\end{eqnarray}
Here, the parameter $h$ is defined above in Eq. (53) by using $C_4$ given in Eq. (65) and here $C_3 = 0$. The PDF tail can now be computed in a similar way as done previously with the result
\begin{eqnarray}
P(\phi_{ZF}) & \sim & \exp\{-\xi_2 (\phi_{ZF}/\phi_{ZF0})^3\}, \\
\xi_2 & = & \frac{4}{27}\frac{1}{h^2} .
\end{eqnarray}
Recall that in the computation of the PDF tails of the zonal flow structure formation, the zonal flow was assumed to be driven by a modon in the ITG turbulence since the bipolar vortex soliton (modon) was assumed to be created by the forcing and that $F(t)=0$ as $t \rightarrow - \infty$, we could treat Eqs (62) and (66) as the transition amplitudes from an initial state with no fluid motion to a state with with different $R/R_0$ or $\phi_{ZF}/\phi_{ZF 0}$. 

Note that all physical quantities are included in the parameters $\xi_1$ and $\xi_2$, through the ion temperature gradient ($\eta_i$), density gradient ($\epsilon_n$), temperature ratio ($\tau=T_i/T_e$), modon size ($a$), modon speed ($U$) and wave number ($k$). It is also important to note that $\xi_j \rightarrow \infty$ (i.e. PDF vanishes) as the the forcing disappears ($\kappa_0 \rightarrow 0$); the instanton cannot form and the PDF vanishes ($P(R) \rightarrow 0$ and $P(\phi_{ZF}) \rightarrow 0$). In addition when the coupling constant $\epsilon \rightarrow 0$ the Reynolds stress vanishes ($R_0 \rightarrow 0$) with vanishing PDF tails ($P(R) \rightarrow 0$ and $P(\phi_{ZF}) \rightarrow 0$).

We note that the exact form of the exponent may depend on the temporal and possibly spatial correlation of the forcing ($f$). In the present paper, the forcing is chosen to be temporally delta correlated for simplicity. The coefficients $\xi_j$ may change if the spatial structure of the solution is changed, i.e. other non-linear solutions to Eq. (1)-(2) were used. In general, for calculating the PDF tail a weighted sum over various coherent structures is needed. At present, the only known exact solution is the modon which we have assumed to be the underlying coherent structure. A generalization should be straightforward if other non-linear solutions were available. The case where the forcing is non-Gaussian ($f$) and other spatial coherent structures will be addressed in future publications.

\section{Results}
We have presented a theory of the PDF tail of structure formation and how the PDF tail of momentum flux is modified by the presence of a shear flow. The exponential forms of the two PDF tails are completely different, signifying the difference in the physical interpretation. In the case of structure formation the PDF tails are found as $\sim \exp\{ -\xi_2 \phi_{ZF}^3\}$, while the momentum flux PDF tail $\sim \exp\{ -\xi_1 R^{3/2}\}$. In this section the parametric dependencies of $\xi_j$ will be studied in detail. The results will be compared with a Gaussian prediction and the results in the case of Hasegawa-Mima (HM) turbulence.

First the PDF tail of momentum flux in ITG turbulence incorporating the effects of shear flow is shown in Figure 1. The parameters are $\eta_i = 4.0 $, $\tau = 0.5$, $\epsilon_n = 1.0$, $g_i = 1$, $a = 2$, $U= 2.0$, $\kappa_0 = 0.3$, $\epsilon = 0.1$, $k \approx 1.91$ with $V_0 =0.0 $ (blue line, solid line), $V_0 = 4.0$ (red line, dash-dotted line), $V_0 = 8.0$ (black line, dotted line),  $V_0 = 12.0$ (green line, dashed line) and $V_0 = 16.0$ (magenta line, thick solid line). When $V_0 = 0$, the result recovers the previous finding in Ref.~\cite{a42}. It is clearly shown that the PDF tail of momentum flux is significantly reduced if a strong shear flow is present whereas weak flow can increase the PDF tail. In the equations (c.f. Eq. (18)-(26)) the flow speed and the modon velocity comes in as a combination of the form ($U-V_0$), determining the behavior of the resulting PDF tails. When ($U-V_0$) decreases, the PDF tail increases until it eventually decreases. This means that there exists a negative value ($U-V_0$) that gives maximum PDF tail, depending on all other parameters. Although due to the Galilean invariance of Eq. (1)-(2) we may always perform Galilean transformations on the instanton solution to find other instanton solutions, those solutions correspond to different ground states in the field theory we consider and must be discarded since they have nonzero velocity in the infinity past~\cite{a70}.

\begin{figure}
  \includegraphics[height=.3\textheight]{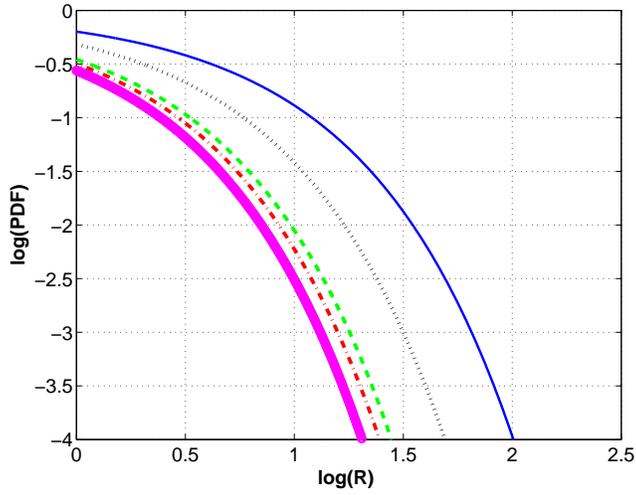}
  \caption{(Color online). The momentum flux PDF tail in ITG turbulence with the effects of shear flow.  The parameters are $\eta_i = 4.0 $, $\tau = 0.5$, $\epsilon_n = 1.0$, $g_i = 1$, $a = 2$, $U= 2.0$, $\kappa_0 = 3 \times 10^{3}$, $\epsilon = 0.1$, $k \approx 1.91$ with $V_0 =0.0 $ (blue line, solid line), $V_0 = 4.0$ (red line, dash-dotted line), $V_0 = 8.0$ (black line, dotted line),  $V_0 = 12.0$ (green line, dashed line) and $V_0 = 16.0$ (magenta line, thick solid line).}
\end{figure}
 
Second, the PDF tails of structure formation as a function of zonal flow potential ($\phi_{ZF}$) are shown in Figure 2. The parameters are $\eta_i = 4.0 $, $\tau = 0.5$, $\epsilon_n = 1.0$, $g_i = 1$, $a = 2$, $V_0 = 12.0$, $U= 2.0$ ($U = -5.0$ dotted black line and dashed black line), $\kappa_0 = 0.3$, $\epsilon = 0.1$ and $k \approx 1.84$ (ITG case), $k=0.81$ (ITG with reversed modon speed), $k \approx 1.73$ (HM case) and $k=1.56$ (HM with reversed modon speed). The PDF tails of ITG turbulence (blue line, solid line), forced Hasegawa-Mima (HM) turbulence (red line, dash-dotted line), Gaussian distribution with same parameters as for the ITG mode turbulence (green line, thick solid line) and cases with negative modon speed in ITG mode turbulence (dotted black line) and in HM turbulence (dashed black line) are displayed. The PDF tail in the HM model is obtained by setting the parameters $\beta=1$ and $\epsilon_n = 0.0$. This is equivalent to letting $\eta_i =0.0$ and $\tau=T_i/T_e = 0.0$. The PDF tails of structure formation in ITG turbulence is enhanced over the HM result indicating that stronger zonal flows are generated in ITG turbulence compared to HM turbulence. This has also been seen in simulations. Note however that the Gaussian prediction ($\sim \exp\{ -\xi_2 \phi_{ZF}^2\}$) is larger than the PDFs of structure formation ($\sim \exp\{ -\xi_2 \phi_{ZF}^3\}$) mainly due to the difference in exponential ($\phi_{ZF}^3 \rightarrow \phi_{ZF}^2$). A reversed modon speed may enhance the PDF tail.
 
\begin{figure}
  \includegraphics[height=.3\textheight]{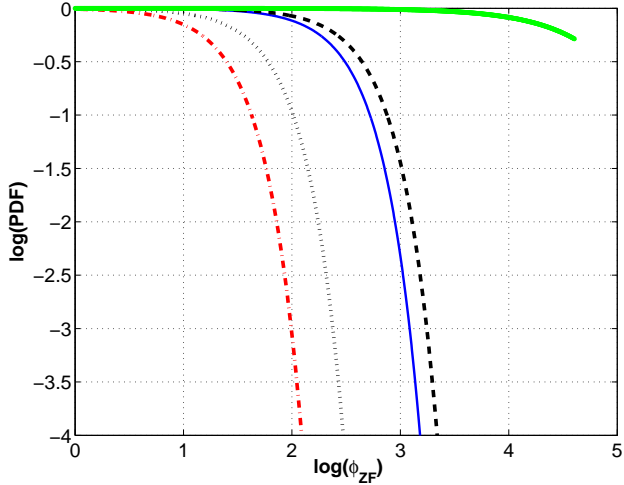}
  \caption{(Color online). The PDF tail of structure formation as a function of zonal flow potential. The PDF tails ITG turbulence (blue line, solid line), forced Hasegawa-Mima (HM) turbulence (red line, dash-dotted line), Gaussian distribution with same parameters as for the ITG mode turbulence (green line, thick solid line) and cases with negative modon speed in ITG mode turbulence (dotted black line) and in HM turbulence (dashed black line). The parameters are $\eta_i = 4.0 $, $\tau = 0.5$, $\epsilon_n = 1.0$, $g_i = 1$, $a = 2$, $V_0 = 12.0$, $U= 2.0$ ($U = -5.0$ dotted black line and dashed black line), $\kappa_0 = 3 \times 10^{3}$, $\epsilon = 0.1$ and $k \approx 1.84$ (ITG case), $k=0.81$ (ITG with reversed modon speed), $k \approx 1.73$ (HM case) and $k=1.56$ (HM with reversed modon speed).}
\end{figure}

Third, the PDF tail of structure formation as a function of zonal flow potential ($\phi_{ZF}$) with shear flow strength ($V_0$) as a parameter is displayed in Figure 3. The resulting PDF tails are shown for $V_0 =0.0 $ (blue line, solid line), $V_0 = 4.0$ (red line, dash-dotted line), $V_0 = 8.0$ (black line, dotted line),  $V_0 = 12.0$ (green line, dashed line) and $V_0 = 16.0$ (magenta line, thick solid line). The other parameters are the same as those in Figure 1. It is found that the PDF tail of structure formation is decreased for large shear flow velocity $V_0$, whereas weak flow can increase the PDF tail. The main reason for this behavior is the same as explained in Figure 1. 

\begin{figure}
  \includegraphics[height=.3\textheight]{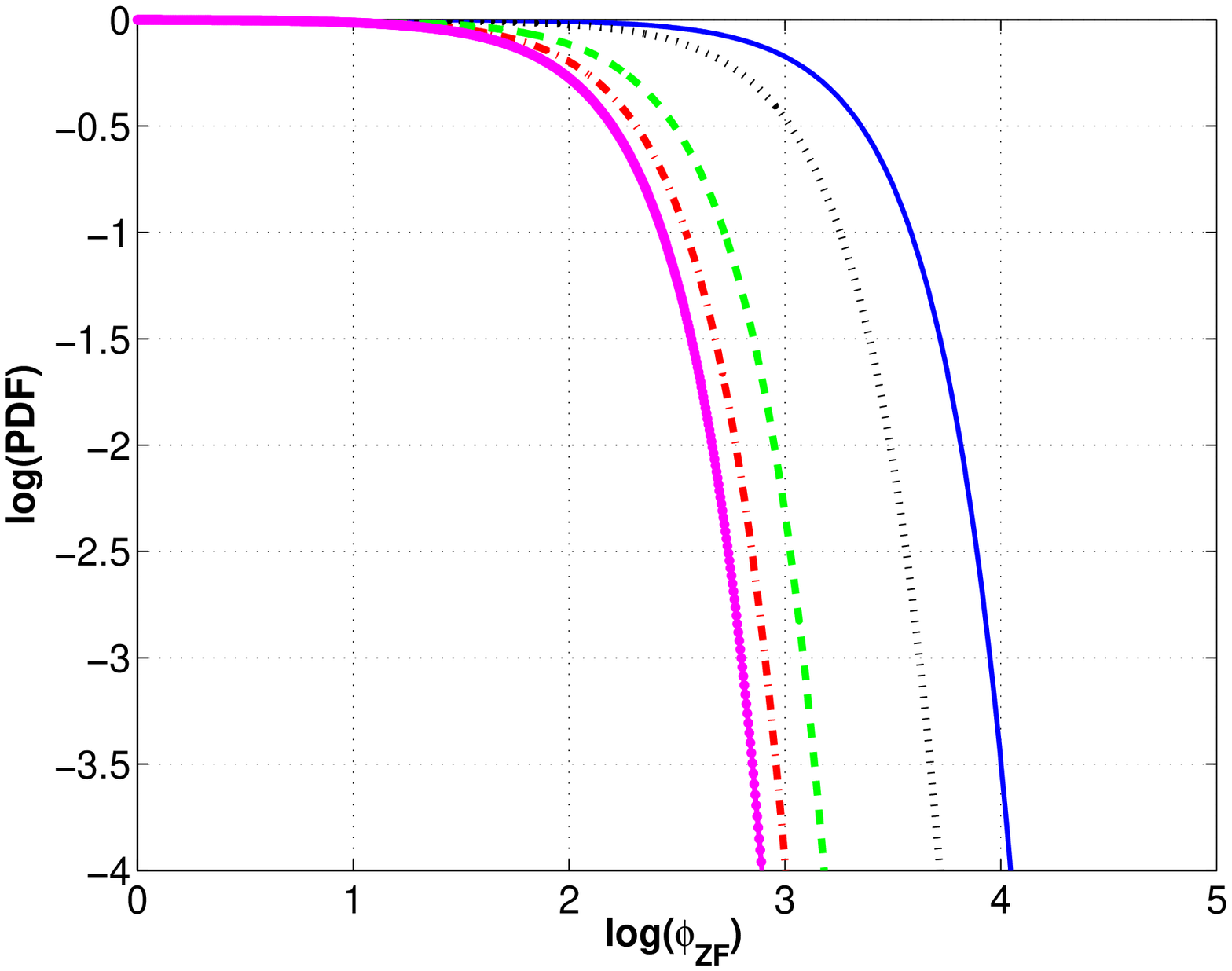}
  \caption{(Color online). The PDF tail of structure formation as a function of zonal flow potential with zonal flow strength as a paramter. The resulting PDF tails are shown for $V_0 =0.0 $ (blue line, solid line), $V_0 = 4.0$ (red line, dash-dotted line), $V_0 = 8.0$ (black line, dotted line),  $V_0 = 12.0$ (green line, dashed line) and $V_0 = 16.0$ (magenta line, thick solid line). The other parameters are the same as in Figure 1.}
\end{figure}

Fourth, the PDF tail of structure formation as a function of zonal flow potential ($\phi_{ZF}$) with $\eta_i$ as a parameter is displayed in Figure 4. The PDF tails are shown for $\eta_i = 2.0$ (red line, dashed line), $\eta_i = 4.0$  (blue line, solid line) and $\eta_i = 6.0$ (black line, dash-dotted line). The parameters are $V_0 = 14.0$, $k \approx 1.91$, $U=4.0$ and the others are the same as in Figure 2. The PDF tail is significantly increased with increasing $\eta_i$, indicating that a stronger zonal flow is generated with increasing normalized temperature gradient $\eta_i$. This suggests that further from marginal stability with a higher level of heat flux, the micro-scale ITG turbulence is more likely to generate stronger zonal flows, leading to a self-regulating system. That is, turbulence and zonal flows constitute a self-regulating system whereby zonal flows generated by turbulence damp turbulence

\begin{figure}
  \includegraphics[height=.3\textheight]{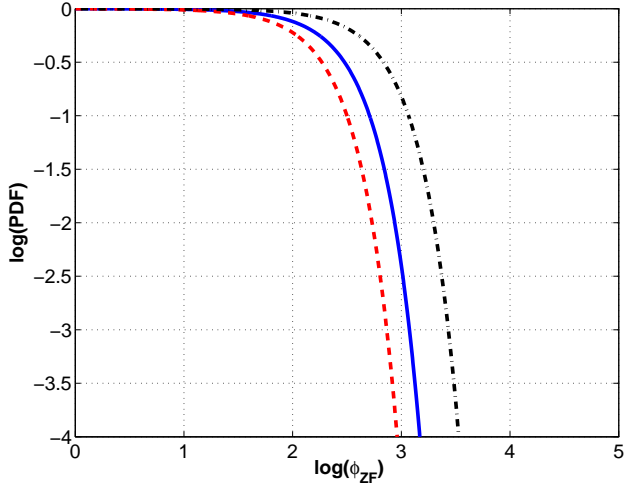}
  \caption{(Color online). The PDF tail of structure formation as a function of zonal flow potential with $\eta_i$ as a parameter. The PDF tails are shown for $\eta_i = 2.0$ (red line, dashed line), $\eta_i = 4.0$  (blue line, solid line) and $\eta_i = 6.0$ (black line, dash-dotted line). The parameters are $V_0 = 14.0$, $k \approx 1.91$, $U=4.0$ and the others are as in Figure 2.}
\end{figure}

\section{Discussion and conclusions}
The PDF tail of momentum flux $\sim \exp\{ -\xi_1 (R/R_0)^{3/2}\}$ has been obtained in previous studies and seems to be ubiquitous in drift wave turbulence~\cite{a19}-~\cite{a42}. This scaling holds when the interaction with shear flow is incorporated as shown in this paper. The origin of this scaling is the quadratic non-linearity in the dynamical system (Eqs 1-2). The difference in scaling of the PDF tails momentum flux and structure formation comes from the change in the temporal behavior of the modon (a change in the initial condition $F_0$). The spatial structure of the modon is of less importance in determining the PDF tails than the temporal behavior. Therefore an approximate spatial structure is sufficient to determine the exponential scaling whereas the time dependency may affect the scaling. Note that the results for momentum flux PDF tails are also applicable to the PDF tails of heat flux, as shown in Ref.~\cite{a42}. The spatial structure of the flow is incorporated in the initial condition of the flow $\phi_{ZF 0}$.

We note that a non-Gaussian scaling of the PDF (the exponent of R) is found even when the forcing is Gaussian, although the exact exponent may depend on the temporal and possibly spatial correlation of the forcing ($f$). In the present paper, the forcing is chosen to be temporally delta correlated for simplicity. The exponent may also change, if another spatial coherent structure is introduced i.e. another non-linear solution to the Eq. (1)-(2) is found. The case where the forcing is non-Gaussian ($f$) will be addressed in a future publication.

In summary, this paper presents the first prediction of the PDF tails of structure formation. One of the important results is that the PDF tail of structure formation from ITG turbulence is significantly increased compared to that in HM turbulence. Zonal flows are also shown to be more likely to be generated further from marginal stability, which will then regulate ITG turbulence, leading to a self-regulating system. Namely, while ITG turbulence is a state with high level of heat flux, it also generates stronger zonal flows that inhibit transport. This also suggests that stronger zonal flows are generated in ITG turbulence compared with ETG turbulence. It was also shown that shear flows can significantly reduce the PDF tails of Reynolds stress and zonal flow formation. 
\section{Acknowledgment}
This research was supported by the Engineering and Physical Sciences Research Council (EPSRC) EP/D064317/1.


\newpage
\begin{thebibliography}{200}
\bibitem{a10} W. Horton, Rev. Mod. Phys. {\bf 71}, 735 (1999)
\bibitem{a35} E. Kim and P. H. Diamond, Phys. Plasmas {\bf 11}, L77 (2004)
\bibitem{a36} E. Kim, Phys. Plasmas {\bf 12}, 090902 (2005)
\bibitem{a51} P. H. Diamond, S-I. Itoh, K. Itoh and T. S. Hahm, Plasma Phys. Contr. Fusion {\bf 47} R35 (2005)
\bibitem{a52} A. Hasegawa, C. G. Mcclennan and Y. Kodama, Phys. Fluids {\bf 22}, 2122 (1979)
\bibitem{a53} H. Biglari, P.H. Diamond and P.W. Terry, Phys. Fluids B {\bf 2}, 1 (1990)
\bibitem{a14} P. A. Politzer, Phys. Rev. Lett. {\bf 84}, 1192 (2000)
\bibitem{a15} P. Beyer, S. Benkadda, X. Garbet and P. H. Diamond, Phys. Rev. Lett. {\bf 85}, 4892 (2000)
\bibitem{a16} J. F. Drake, P. N. Guzdar and A. B. Hassam, Phys. Rev. Lett. {\bf 68}, 2205 (1988)
\bibitem{a38} B. A. Carreras, B. van Milligen, C. Hidalgo R. Balbin, E.
Sanchez, I. Garcia-Cortez, M. A. Pedrosa, J. Bleuel and M. Endler, Phys. Rev. Lett. {\bf 83}, 3653 (1999)
\bibitem{a19} E. Kim and P. H. Diamond, Phys. Plasmas {\bf 9}, 71 (2002)
\bibitem{a20} E. Kim and P. H. Diamond, Phys. Rev. Lett. {\bf 88}, 225002 (2002)
\bibitem{a21} E. Kim, P. H. Diamond, M. Malkov, T.S. Hahm, K. Itoh, S.-I. Itoh, S. Champeaux, I. Gruzinov, O. Gurcan, C. Holland, M.N. Rosenbluth and A. Smolyakov, Nucl. Fusion {\bf 43}, 961 (2003)
\bibitem{a81} W. Horton and Y.-H. Ichikawa,  Chaos and Structures in Nonlinear Plasmas (World Scientific 1996) 238 
\bibitem{a42} J. Anderson and E. Kim, Phys. Plasmas, {\bf 15} 052306 (2008)
\bibitem{a13} J. Anderson, H. Nordman, R. Singh and J. Weiland, Phys. Plasmas {\bf 9}, 4500 (2002)
\bibitem{a25} V. D. Larichev and G. M. Reznik, Dokl. Akad. Nauk SSSR {\bf 231}, 1077 (1976)
\bibitem{a26} B. G. Hong, F. Romanelli and M. Ottaviani, Phys. Fluids B {\bf 3}, 615 (1991) 
\bibitem{a27} F. L . Waelbroeck, P. J. Morrison and W. Horton Plasma Phys. and Contr. Fusion {\bf 46}, 1331 (2004)
\bibitem{a11} G. Bateman, A. H. Kritz, J. E. Kinsey, A. Redd and J. Weiland,, Phys. Plasmas {\bf 5} 1793 (1998)
\bibitem{a12} A. M. Dimits, G. Bateman, M. A. Beer, B. I. Cohen, W. Dorland, G.W. Hammett, M. A. Beer, B. I. Cohen, C. Kim, J. E. Kinsey, M. Kotschenreuther, A. H. Kritz, L. L. Lao, J. Mandrekas, W. M. Nevins, S. E. Parker, A. J. Redd, D. E. Shumaker, R. Sydora and J. Weiland, Phys. Plasmas {\bf 7} 969 (2000)
\bibitem{a22} J. Zinn-Justin, Field Theory and Critical Phenomena (Clarendon, Oxford, 1989)
\bibitem{a37} J. Weiland, Collective Modes in Inhomogeneous Plasmas, Kinetic and Advanced Fluid Theory (IOP Publishing Bristol 2000) 115
\bibitem{a61} F. Jenko, W. Dorland, M. Kotschenreuther and B. N. Rogers, Phys. Plasmas {\bf 7}, 1904 (2000)
\bibitem{a28} J. Anderson, H. Nordman and J. Weiland, Plasma Phys. Contr. Fusion {\bf 42}, 545 (2000)
\bibitem{a29} J. Anderson, H. Nordman and J. Weiland, Phys. Plasmas {\bf 8}, 180 (2001)
\bibitem{a70} V. Gurarie and A. Migdal, Phys. Rev. E {\bf 54}, 4908 (1996) 
\end{thebibliography}
\end{document}